\documentclass[reprint,amsmath,amssymb,longbibliography,aps, prb, floatfix]{revtex4-2}
\pdfoutput=1
\usepackage{hyperref}
\usepackage{url}
\usepackage{color}
\usepackage{acro}
\usepackage{chemmacros}
\usepackage[normalem]{ulem}
\usepackage{graphicx}
\usepackage[a4paper, total={7in, 10in}]{geometry}
\usepackage{amsmath}
\usepackage{soul}

\begin{document}

\title{Anisotropy lifts topological protection of transport in pumped spin chains
}
\author{Mainak Pal$^{1}$\thanks{email for correspondence: mainak.pal@ufl.edu}
}

\affiliation{$^1$Department of Physics and the Quantum Theory Project, University of Florida, Gainesville, Florida, USA}

\begin{abstract}
We investigate the prospect of topological spin {pumping and transport} in a dimerized spin-$\frac{1}{2}$ anisotropic XY chain with tunable in-plane anisotropy. Through a Jordan–Wigner transformation, the system maps onto an extended Rice–Mele-type fermionic model whose momentum-space band structure exhibits gapped regions characterized by nonzero Chern numbers ($\pm 1$) as well as trivial and gapless regimes. Conventional momentum-space analysis therefore predicts quantized Thouless-type spin {pumping} in parts of the Chern number diagram. However, real-space time evolution under open boundary conditions reveals that the  spin {transport} is generally not quantized away from the isotropic limit. In the anisotropic regime, eigen-energy states within the {transporting} subspace hybridize during the driving cycle, leading to a transport that depends on details of the driving protocol. Quantized pumping and transport survive only in the isotropic limit, where the Hamiltonian reduces to decoupled Rice–Mele sectors and the transported spin component is conserved. Our results demonstrate that momentum-space topological invariants alone do not guarantee quantized spin transport in anisotropic spin chains. More broadly, they show that the conventional Thouless-pumping framework does not straightforwardly extend to situations where the transported quantity is not conserved, highlighting the importance of real-space diagnostics in topological transport phenomena.
\end{abstract}
\maketitle

\section{Introduction}
Topological pumping, originally proposed by Thouless in one-dimensional fermionic systems \cite{Thouless1983, Niu_1984}, is a paradigmatic example of quantized charge transport protected by topology. In a Thouless pump, adiabatic and cyclic modulation of system parameters transports an integer number of particles across the system during each driving cycle, with the transported quantity determined by a topological invariant of the occupied bands. The Rice–Mele model \cite{RiceMele,Asb_th_2016} provides the canonical realization of this phenomenon and has motivated extensive theoretical and experimental exploration of topological transport in condensed matter and synthetic quantum systems \cite{bastidas2022,Barouch1970,Barouch1971,Lohse2016,Nakajima2016,Lohse2018,Kraus2012,verbin2015,Zilberberg2018,Stefan2014,Fu-Sheng2018,godfrin2017,Liddle2015,Hussain2018,Chicco2021,Liu2021,Tacchino2021}.

In recent years, there has been growing interest in extending these ideas from fermionic systems to spin systems \cite{bastidas2022,Barouch1970,Barouch1971}. Modern experimental platforms, including cold atoms \cite{Lohse2016,Nakajima2016,Lohse2018}, waveguide arrays \cite{Kraus2012,verbin2015,Zilberberg2018}, and magnetic molecular complexes \cite{Stefan2014,Fu-Sheng2018,godfrin2017,Liddle2015,Hussain2018,Chicco2021,Liu2021,Tacchino2021}, now permit increasingly precise control over effective spin-chain Hamiltonians and time-dependent couplings. Such advances naturally raise the question of whether topologically protected Thouless pumping can be realized robustly in spin systems.

{Two particularly important settings of quantum spin chains are provided by the isotropic XX model and the quantum Ising (QI) model, two paradigmatic and exactly solvable limits of spin-1/2 systems. The XY chain continuously interpolates between these limits through an in-plane anisotropy parameter $\alpha$, which controls the relative strength of the $x$- and $y$-components of the exchange interactions ($\alpha=0$ corresponds to the XX limit, while $\alpha=1$ gives the QI limit) \cite{TONG200191, LIEB1961407, Young1996, Fisher1992, Doria1988, Benza_1989, Luck1993, Luck_1986, Satija1989, Juozapavi1999, Barouch1970, Barouch1971, Derzhko2004, Derzhko_2002, Çakmak2015, Batle2010, Hoffman2023}. In the isotropic XX limit, the rotational symmetry $SO(2)$ about the $z$-axis is preserved, the $z$-component of spin is conserved, and the Jordan--Wigner transformation maps the spin-chain directly onto the Rice--Mele model, quantized spin pumping and transport follow straightforwardly from the corresponding fermionic topology. Away from the isotropic limit, however, the anisotropy breaks this symmetry and generates pairing terms in the fermionic representation that mix particle and hole sectors. Consequently, once the $z$-component of spin is no longer conserved, it becomes unclear how the transport is modified, and the correspondence between momentum-space topology and physical spin transport becomes less transparent.} 

{Here we investigate the prospect of topological spin pumping and transport in a dimerized anisotropic XY chain with periodically modulated exchange interactions and staggered magnetic fields. We first construct the momentum-space Chern-number diagram in the parameter space of average intra-dimer coupling and spin anisotropy, identifying gapped regions with nontrivial Chern numbers that would conventionally be associated with quantized pumping. We then test these predictions through real-space simulations under open boundary conditions and find that, although nontrivial Chern numbers persist over a broad parameter range, quantized spin pumping survives only in the isotropic limit. In the anisotropic regime, hybridization within the transporting subspace leads to protocol-dependent, non-quantized transport, a breakdown further exposed by a modified wait-time protocol (that includes a pause in parameter variation during the wait-time) and traced to the loss of spin conservation.}

The rest of the paper is organized as follows. Section II introduces the anisotropic dimerized XY chain, the Jordan–Wigner mapping, the pumping protocol, and the expression of the topological quantum number associated with the relevant energy bands. Section III first presents the momentum-space topological characterization and the resulting Chern-number diagram as a function of anisotropy and average intra-dimer exchange. Both the isotropic and anisotropic regimes are discussed. Next, effect of a small higher-frequency perturbation in the pumping protocol is explored. Section IV investigates real-space dynamics and demonstrates the breakdown of quantized transport away from the isotropic limit. Section V discusses the physical origin of this breakdown. Section VI summarizes our work and discusses its implications for experiments related to transport in spin systems.

\section{Spin chain model}

\subsection{{Spin Hamiltonian and system parameters}}
We consider a dimerized spin-$\frac{1}{2}$ anisotropic XY chain with staggered nearest-neighbor exchange interactions. The Hamiltonian of the chain is

\begin{equation} \label{HXY}
\begin{split}
H_{\mathrm{XY}}
&= \dfrac{(1+\alpha)}{2}\Bigg\{\sum_{i=1}^{N-1}\Big(\ v\ \sigma_{2i-1}^{x}\sigma_{2i}^{x} + w\ \sigma_{2i}^{x}\sigma_{2i+1}^{x}\Big)
\\ &+ \ v\ \sigma_{2N-1}^{x}\sigma_{2N}^{x} + \eta \ w\ \sigma_{2N}^{x}\sigma_{1}^{x}\Bigg\}
\\ &+ \dfrac{(1-\alpha)}{2}\Bigg\{\sum_{i=1}^{N-1}\Big(\ v\ \sigma_{2i-1}^{y}\sigma_{2i}^{y} + w\ \sigma_{2i}^{y}\sigma_{2i+1}^{y}\Big)
\\ &+ \ v\ \sigma_{2N-1}^{y}\sigma_{2N}^{y} + \eta \ w\ \sigma_{2N}^{y}\sigma_{1}^{y}\Bigg\}
\end{split}
\end{equation}
where $v$ and $w$ denote the intra-dimer and inter-dimer exchange interactions respectively, $i$ is the dimer index, $N$ is the number of dimers, and $\sigma^{x,y,z}$ are the spin operators.
{When the first and last dimers are not connected to each other, we have open boundary condition (OBC), and when they are connected via an inter-dimer exchange interaction, we have periodic boundary condition (PBC). The variable $\eta$ takes two discrete values and represents either OBC ($\eta = 0$) or PBC ($\eta = 1$).} The parameter $\alpha$ controls the in-plane anisotropy of the exchange interaction.
{The anisotropy parameter takes value continuously in the range $[0,1]$. The isotropic limit $\alpha=0$ corresponds to the conventional XX model, while $\alpha=1$ reduces to the QI limit.}

\begin{figure}[tb]
\centering
\includegraphics[width=1\linewidth]{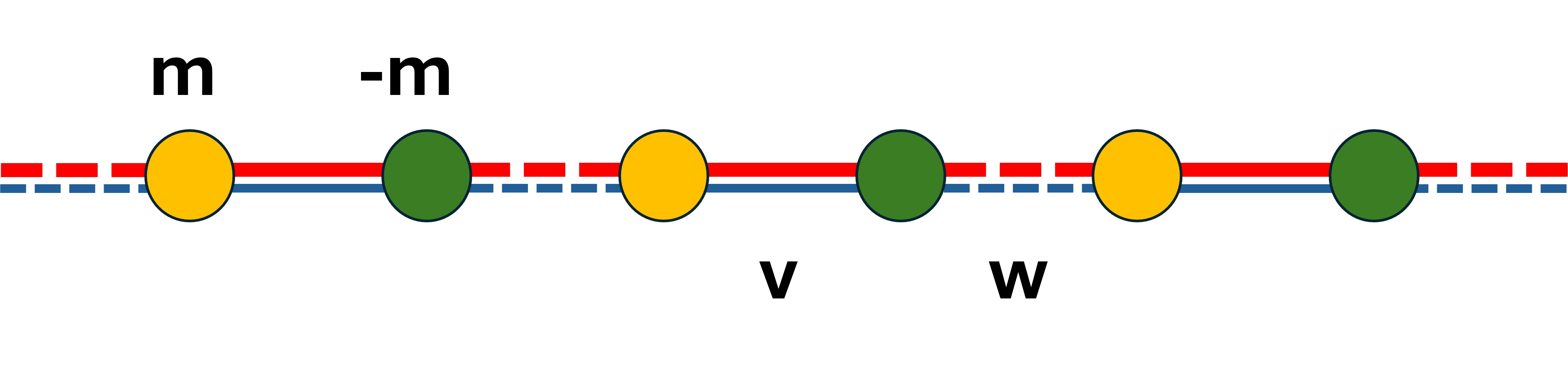}    \caption{Schematic of the dimerized anisotropic XY chain and the relevant parameters.
Solid and dashed bonds denote intra-dimer and inter-dimer exchange interactions $v$ and $w$, respectively. Red and blue bonds represent the $x$- and $y$-components of the exchange interaction, whose relative strength is controlled by the anisotropy parameter $\alpha$. Yellow and green sites indicate the two sub-lattices with alternating signs of the staggered magnetic field $m$.}
\label{fig:ExtendedRiceMeleSchematic}
\end{figure}

Additionally a staggered transverse magnetic field is introduced,
\begin{equation} \label{HB}
\begin{split}
H_{\mathrm{B}}
&= m\sum_{i=1}^{N}\Big(\sigma_{2i-1}^{z}-\sigma_{2i}^{z}\Big)
\end{split}
\end{equation}
where $m$ controls the strength of the field.
Fig. \ref{fig:ExtendedRiceMeleSchematic} schematically illustrates the system and the relevant parameters. The staggered exchange interactions and the alternating magnetic field differentiate the two sub-lattices. 
Throughout this work, periodic boundary conditions are employed for momentum-space calculations, whereas open boundary conditions are used for real-space transport simulations.

\subsection{Spin fermion map and fermionized Hamiltonian}
We map the XY spin model
to  the fermion model of Rice-Mele via a Jordan-Wigner transformation of the spin operators to their fermion counterpart \cite{Coleman_2015}. The spin operators $\sigma$ are expressed as
\begin{equation*}
    \begin{split}
        &\sigma_{2j-1}^{z} = a_{j}^{\dagger}a_{j}-1/2,\ \ \sigma_{2j}^{z} = b_{j}^{\dagger}b_{j}-1/2
    \end{split}
\end{equation*}
\begin{equation*}
    \begin{split}
        &\sigma_{2j-1}^{+} = a_{j}^{\dagger}e^{i\pi\sum_{l<j}(n^{a}_{l}+n^{b}_{l})}
    \end{split}
\end{equation*}
\begin{equation}
    \begin{split}
        &\sigma_{2j}^{+} = b_{j}^{\dagger}e^{i\pi\sum_{l<j}\{(n^{a}_{l}+n^{b}_{l})+n^{a}_{j}\}}
    \end{split}
\end{equation}
where $a_{j}^{\dagger},\ b_{j}^{\dagger}$ are the fermion creation operators for the two sub-lattice sites of the $j^{\rm th}$ dimer, and $n^{a,b}_{l}$ are respectively the fermion occupancy at the $l^{\rm th}$ dimer for the two sub-lattice sites.
The pumped charge in the fermion picture corresponds to the spin component parallel to the $z$ axis in the spin picture. The spin component anti-parallel to the $z$ axis corresponds to holes. After fermionization, the Hamiltonian in eq. \ref{HXY} reads
\begin{equation}
\begin{split}
H_{\rm{XY}}&=\dfrac{(1+\alpha)}{2}\Bigg\{\frac{v}{4}\sum_{i=1}^{N}\Big(a_{i}^{\dagger}b_{i}^{\dagger} - a_{i}b_{i} + a_{i}^{\dagger}b_{i} - a_{i}b_{i}^{\dagger}\Big) 
\\&+ \frac{w}{4}\sum_{i=1}^{N-1}\Big(b_{i}^{\dagger}a_{i+1}^{\dagger} - b_{i}a_{i+1} + b_{i}^{\dagger}a_{i+1} - b_{i}a_{i+1}^{\dagger}\Big)
\\&+ \eta \frac{w}{4}\Big(b_{N}^{\dagger}a_{1}^{\dagger} - b_{N}a_{1} + b_{N}^{\dagger}a_{1} - b_{N}a_{1}^{\dagger}\Big)\Bigg\}\\
&+\dfrac{(1-\alpha)}{2}\Bigg\{\frac{v}{4}\sum_{i=1}^{N}\Big(-a_{i}^{\dagger}b_{i}^{\dagger} + a_{i}b_{i} + a_{i}^{\dagger}b_{i} - a_{i}b_{i}^{\dagger}\Big) 
\\&+ \frac{w}{4}\sum_{i=1}^{N-1}\Big(-b_{i}^{\dagger}a_{i+1}^{\dagger} + b_{i}a_{i+1} + b_{i}^{\dagger}a_{i+1} - b_{i}a_{i+1}^{\dagger}\Big)
\\&+ \eta \frac{w}{4}\Big(-b_{N}^{\dagger}a_{1}^{\dagger} + b_{N}a_{1} + b_{N}^{\dagger}a_{1} - b_{N}a_{1}^{\dagger}\Big)\Bigg\}
\end{split}
\end{equation}
where $a_{i}^{\dagger}$, $b_{i}^{\dagger}$ are the creation operators at dimer $i$ for the two sub-lattices, and $a_{i}$, $b_{i}$ are the corresponding annihilation operators.
The full Hamiltonian combining eqs. \ref{HXY} and \ref{HB} in the momentum space takes the form (with PBC, i.e. $\eta = 1$) as
\begin{equation}
\label{eq:HamXYB}
\begin{split}
    \hat{H}_{\mathrm{XY,B}}
&=
\sum_{k} \Psi_{k}^{\dagger}H_{k}\Psi_{k}
\end{split}
\end{equation}
with 
$\Psi_{k} = \big[
    a_{k}, \ b_{k}, \ a_{-k}^{\dagger}, \ b_{-k}^{\dagger}\big]^{T}$ and
\begin{widetext}
\begin{equation*}
    \begin{split}
        H_{k} = \begin{bmatrix}
    m & v+w\ e^{-ik} & 0 & \alpha(v-w\ e^{-ik})\\
    v+w\ e^{ik} & -m & \alpha(-v+w\ e^{ik}) & 0\\
    0 & \alpha(-v+w\ e^{-ik}) & -m & -v-w\ e^{-ik}\\
    \alpha(v-w\ e^{ik}) & 0 & -v-w\ e^{ik} & m
\end{bmatrix}
    \end{split}
\end{equation*}
\end{widetext}
where $a_{k}^{\dagger}$, $b_{k}^{\dagger}$ are the creation operators at momentum $k$ for the two sub-lattices, and $a_{k}$, $b_{k}$ are the corresponding annihilation operators. Thus, the XY spin-Hamiltonian naturally adopts the Nambu form, as expressed using the momentum space fermion operators.

{In the isotropic limit $\alpha=0$, the off-diagonal pairing blocks vanish.  
In this case the transported spin component is conserved due to the preserved $SO(2)$ symmetry, and, as we'll show, the conventional quantized Thouless pumping and transport continues. For nonzero anisotropy, however, the pairing terms hybridize the particle and hole sectors, reflecting the lack of rotational symmetry in the spin space, which will manifest in imperfect transport.}  

\subsection{Pumping protocol}
The Hamiltonian introduced in the last subsection describes a static chain with a specific anisotropy and dimerization in the in-plane spin exchange interaction. To drive the chain, the spin-exchange interactions as well as the onsite magnetic fields need to be dynamically varied as a function of time, such that a difference between the phases of their time variation is maintained. By that, the time-dependent configuration of up and down magnetic fields at alternate sites can be intuitively understood to slide on the chain like a wave, thereby guiding the spin transport. Modulation of the interactions and fields can be intuitively understood as operating valves or one-way doors to assist the process. 
It can be achieved with various protocols of temporal variation of the system parameters. In \cite{bastidas2022} a single but both spatially and temporally varying parameter was used. 
However, motivated by Thouless pumping in the Rice-Mele model, we adopt a simple choice, following \cite{Asb_th_2016}, for the temporal variation of the system variables in this work as follows,
\begin{equation} \label{eq:time_protocol}
    \begin{split}
       &m(t)=\mathrm{sin}(\Omega t)\,, \\
       &v(t)=\bar{v}+\mathrm{cos}(\Omega t)\,,\\
       &w=1\,,
    \end{split}
\end{equation}
where $\Omega = 2\pi/T$ is the angular frequency and $T$ is the time-period of the cycle of the variation. Here, $\bar{v}$ denotes the average of the intra-dimer exchange interaction over one cycle. This is the time-dependence of the variables that has been followed in the rest of this article (with the time-period denoted as $T_{\rm{nw}}$) unless explicitly mentioned otherwise, but suppressed in the symbols to avoid clutter. 

\begin{figure*}[tb]
\centering
\includegraphics[width=1\linewidth]{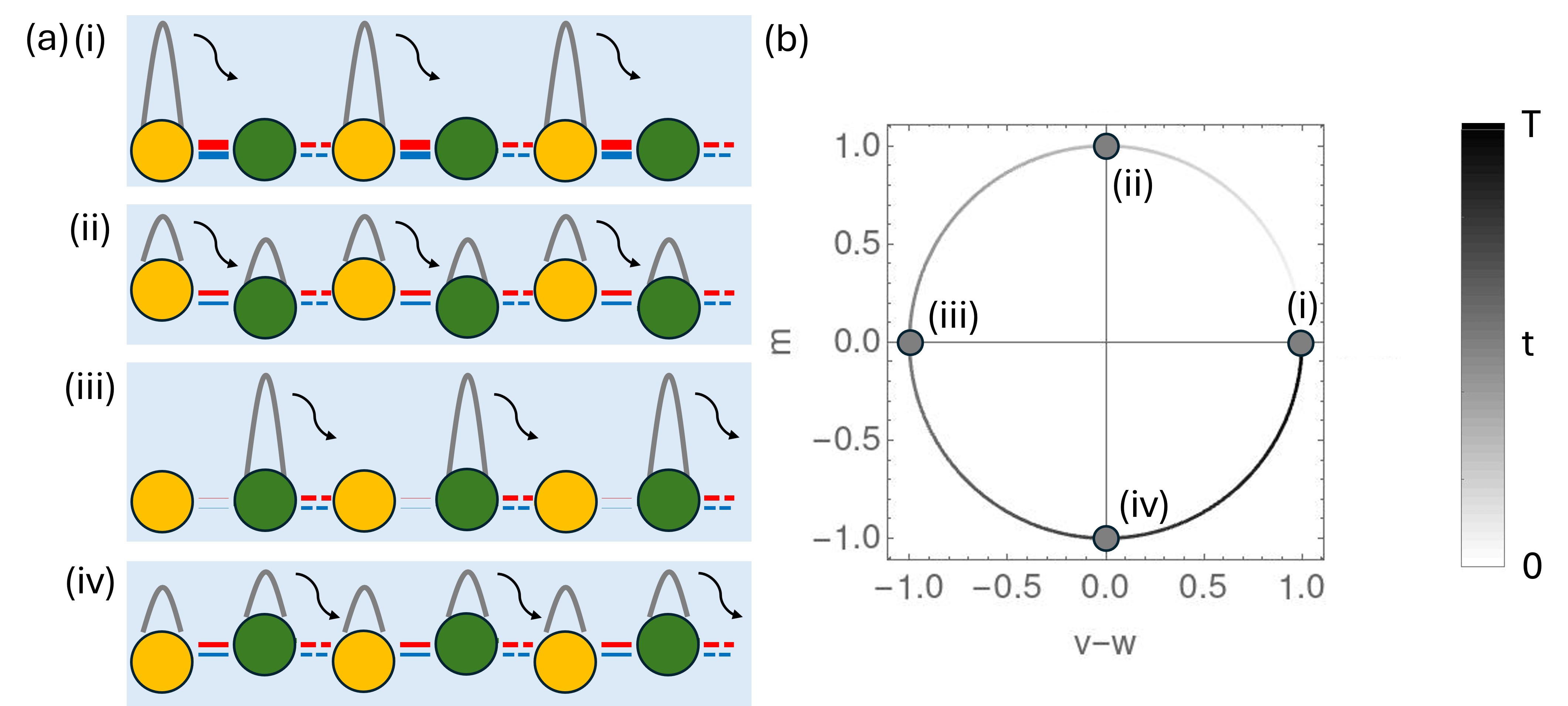}    \caption{Schematic of the time-variation protocol in the chain during one cycle. (a) (i-iv) are sequential snapshots of the chain during a pumping cycle, qualitatively showing the transport of a state from one sub-lattice to the other. 
The height or depth of the two sub-lattices (yellow and green balls) relative to the interaction lines represent the strength of the antiparallel magnetic fields $m$, $-m$ on two sub-lattices at the corresponding instant. Directional pumping of the particle density (gray peaks) initially distributed on one sub-lattice is guided by time variation in the  spin-exchange interaction and differential magnetic field between neighboring sites. (b) $v-w$ vs $m$ parametric plot as the parameter $t$ varies from $0$ to $T$ through the instants (i-iv) as shown in (a).}
\label{fig:vw_vs_m_polished_combined}
\end{figure*}

{Fig. \ref{fig:vw_vs_m_polished_combined}(a) shows the protocol schematically, depicting how the time-variation of the exchange interactions in synchrony with the magnetic fields is expected to accomplish the transport in the chain, with periodic boundary condition assumed.}
{A parametric plot of $v-w$ vs $m$ as a function of time is shown in Fig. \ref{fig:vw_vs_m_polished_combined}(b) for $\bar{v}=1$. Here $v-w$ is a measure of degree of dimerization of the spin exchange interaction in the chain.
}

\subsection{{Condition and expression of quantized pumping}} 
Imposing the periodic boundary condition  to spatially identify the two ends of the chain, along with the already existing periodicity in time, gives us a two-dimensional toroidal space comprising one space and one time coordinates, and allows calculating an instantaneous momentum-resolved pumped current,
{with quantized value of the pumped charge} \cite{Thouless1983, Asb_th_2016}. 

The Hamiltonian in eq. \ref{eq:HamXYB} admits four eigenstates at any momentum and time instant, and therefore forms four energy-bands or energy-surfaces as functions of momentum over the time period of the protocol. We work in the half-filled scenario, i.e. two filled bands.
The time and momentum dependent current in the adiabatic limit ($\Omega\equiv2\pi/T \rightarrow0$, see eq. \ref{eq:time_protocol}) is the Pancharatnam-Berry \cite{Pancharatnam1956,Berry1984} curvature of the filled band i.e.,
\begin{equation}
\label{eq:instcur}
\begin{split}
j(k,t)=-i\big(&\langle\partial_{k}u
(k,t)\vert\partial_{t}u
(k,t)\rangle\\ 
-&\langle\partial_{t}u
(k,t)\vert\partial_{k}u
(k,t)\rangle\big)
\end{split}
\end{equation}
where $\vert u
(k,t)\rangle$ is an instantaneous eigenstate of the Hamiltonian in eq. \ref{eq:HamXYB}.
The total instantaneous current can then be calculated by simply taking the momentum integral over the Brillouin zone in the limit of long chain \cite{RuiLi2017}, 
\begin{equation}
I(t) = \int_{-\pi}^{\pi}\frac{dk}{2\pi}j(k,t)
\end{equation}
Finally, the total pumped charge over one time-period can be obtained by further taking the time-integral over one time period, i.e.
\begin{equation}
    Q=\int_{0}^{T}dt\ I(t)=\int_{0}^{T}dt\ \int_{-\pi}^{\pi}\frac{dk}{2\pi}j(k,t)
\end{equation}
Thus $Q$ is an integral over the closed surface of the torus, and is a topologically quantized quantity as argued in Ref. \cite{Thouless1983} provided the energy band-gap of the periodic Hamiltonian does not close. The conventional Rice-Mele model that corresponds to the isotropic XX spin-chain limit, uses this form of the instantaneous momentum resolved current as in eq. \ref{eq:instcur} which we also use in the anisotropic regime.

In our model we consider two filled bands and each one admits a corresponding $Q$. These pumped charges $Q$ for the two filled bands are the topological quantities of our interest, and are one-dimensional dynamical analogue of the Chern number.
The two bands arise naturally as the Hamiltonian is expressed in the Nambu form, and behave identically, as expected, in the $\alpha = 0$ limit. In the $\alpha \neq 0$ regime they interact and split in energy at times during the pumping protocol. Their average pumping behavior decides the transport in the system. Thus, we characterize the state of our \textit{one space dimensional system over one time period} by the quantity $\mathcal{Q} \equiv (Q_{\rm u} + Q_{\rm l})/2$, where $Q_{\rm u}$ and $Q_{\rm l}$ are the individual values of the topological invariant $Q$ for the upper and lower filled bands respectively, obtained by diagonalizing the Hamiltonian in eq. \ref{eq:HamXYB}.

\section{Chern diagram: $k$-space analysis}
We first analyze the system within the conventional momentum-space framework, where quantized adiabatic transport is inferred from topological invariants associated with the occupied bands. The objective of this section is to identify parameter regimes that, based on band topology alone, would be expected to control Thouless-type quantized spin pumping.
The topology of the pumping cycle is characterized by the total pumped quantity $\mathcal Q$, obtained from the occupied bands over one driving period. We focus primarily on the parameter space spanned by the average intra-dimer coupling $\bar{v}$ and the anisotropy parameter $\alpha$. 

\begin{figure}[tb]
    \centering
\includegraphics[width=1\linewidth]{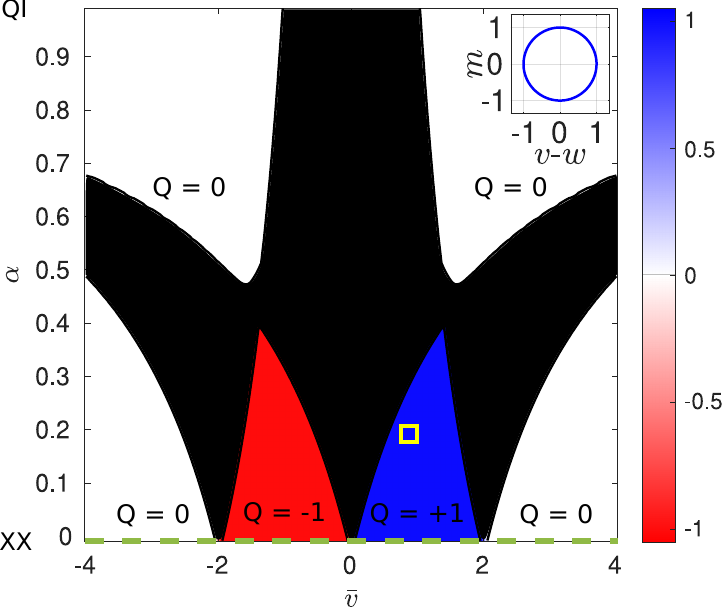}
    \caption{Chern number diagram of the spin-chain over one time period as a function of average intra-dimer exchange $\bar v$ and anisotropy $\alpha$. Blue and red regions correspond to nontrivial topology with $\mathcal Q=\pm1$, respectively. The white region is topologically trivial ($\mathcal Q=0$), while the dark region denotes closing of the instantaneous energy band-gap during the pumping cycle. The inset shows  $m$ vs $v - w$ temporal-drive loop. The green dotted line represents the isotropic limit.
   \label{fig:PhaseDiag}}
\end{figure}

Fig. \ref{fig:PhaseDiag} summarizes the resulting Chern-number diagram based on the calculation described earlier. The diagram contains three qualitatively distinct regimes: gapped regions with nonzero Chern numbers $\mathcal{Q}=\pm1$,
gapped but topologically trivial regions with $\mathcal{Q}=0$, gapless regions where the band topology becomes ill-defined.
{The blue and red regions can be interpreted to correspond to opposite directions of transport based on the conventional momentum space calculation}, while the white region is topologically trivial despite remaining gapped. The dark region indicates closing of the instantaneous band gap during the pumping cycle, preventing a well-defined topological characterization. 

\subsection{The topological isotropic chain
}
\begin{figure}[tb]
\centering
\includegraphics[width=1\linewidth]{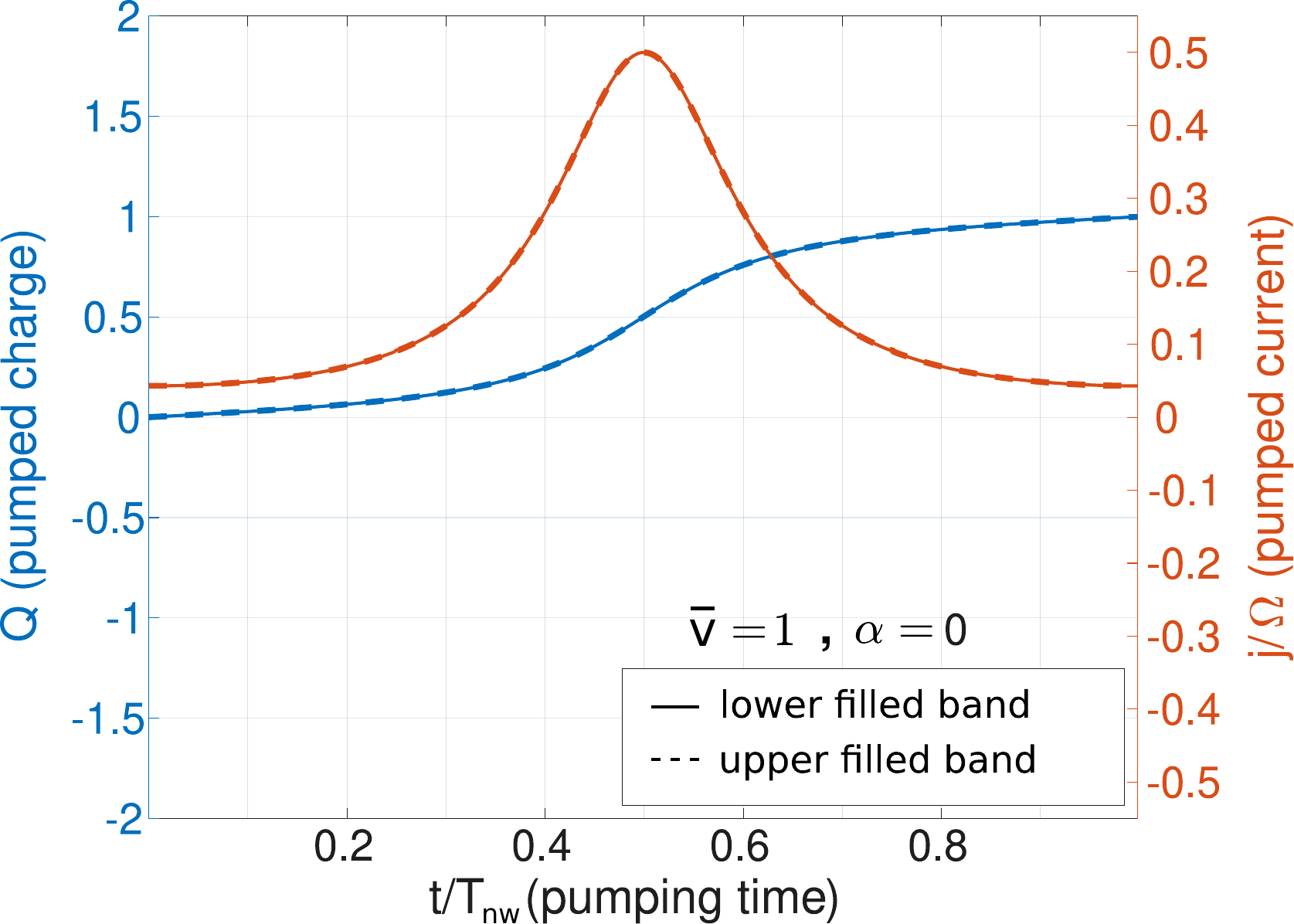}
\caption{Momentum-space pumping characteristics in the isotropic limit $\alpha=0$ for $\bar v=1$. The two occupied bands contribute equally to the pumped quantity, yielding quantized transport ($\mathcal{Q} = 1$) consistent with the Rice–Mele correspondence.}
\label{fig:0only}
\end{figure}

In the Chern diagram we first focus on the well-understood isotropic limit of the system (horizontal green dotted line at $\alpha \rightarrow 0$).
Fig. \ref{fig:0only} shows the momentum-resolved current and pumped charge during one pumping cycle for a typical average intra-dimer exchange $\bar v = 1$ . Each of the upper and the lower filled bands features a topological number of $Q_{\rm u} = Q_{\rm l} = +1$, thus characterizing the isotropic limit of the chain by $\mathcal{Q} \equiv (Q_{\rm u} + Q_{\rm l})/2 = +1$. At $\alpha = 0$ limit, the pumped current and charge in both the filled bands look identical and the momentum-space description reproduces the conventional Thouless-pumping picture. This is as expected, because the fully isotropic XY chain is essentially equivalent to two decoupled mathematical descriptions of the simple Rice-Mele chain, accommodating both particle and hole character, which in the spin picture are essentially $\uparrow$-spin and $\downarrow$-spin. 

As the intra-dimer spin exchange interaction is varied, the isotropic limit of the spin chain is characterized by three values of the quantized Chern number (see Fig. \ref{fig:PhaseDiag} again, $\alpha = 0$ line), namely $\mathcal{Q} = -1$ for $-2 < \bar{v} < 0$, $\mathcal{Q} = +1$ for $0 < \bar{v} < 2$, and $\mathcal{Q} = 0$ for $ \vert \bar{v}\vert > 2$. Importantly, throughout the isotropic regime the system remains gapped during the pumping cycle except $\bar{v} = 0,\pm2$, ensuring a well-defined topological invariant and topological transition points. The isotropic chain therefore behaves exactly as expected from the Rice–Mele correspondence and serves as the benchmark for the anisotropic cases discussed below (see Appendix \ref{appendix:b}).

\subsection{Chern number in anisotropic chain}
Next we focus on the anisotropic chain ($\alpha \neq 0$), where the rotational symmetry $SO(2)$ in the spin picture around the z-axis breaks. 
As the anisotropy increases away from the isotropic limit, the momentum-space analysis continues to predict extended regions of nontrivial topology. 
In particular, the Chern diagram of Fig. \ref{fig:PhaseDiag} contains finite gapped regions with quantized values $\mathcal Q = \pm1$.

\begin{figure}[tb]
    \centering
\includegraphics[width=1\linewidth]{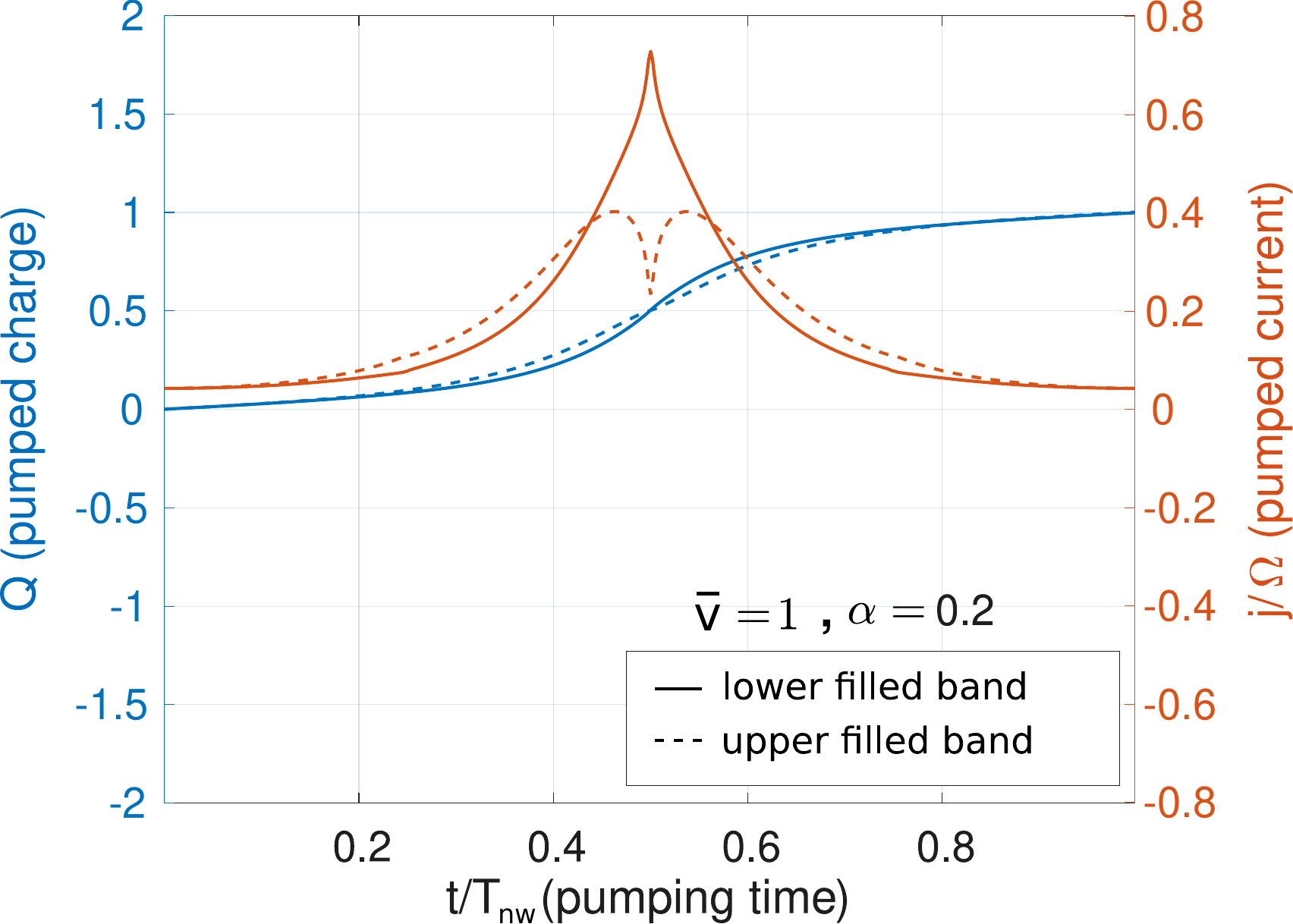}
\caption{Momentum-space pumping characteristics in the anisotropic chain for $\bar v=1$. Moderate anisotropy ($\alpha=0.2$), where the occupied bands contribute additively to the pumping. Details of the current differ but the transport appears quantized ($\mathcal{Q} = 1$).}
\label{fig:ExtendedRiceMeleCurrentCharge2}
\end{figure}

With fixed $\bar v = 1$ and increasing anisotropy $\alpha$, the occupied bands continue to contribute additively to the pumped quantity. Fig. \ref{fig:ExtendedRiceMeleCurrentCharge2} illustrates a representative case with $\alpha = 0.2$, where the two occupied bands still yield $Q_{\rm{u}} = Q_{\rm{l}} = +1$, although the details of instantaneous pumped current corresponding to the occupied bands begin to differ. 

This behavior reflects the increasing influence of anisotropy-induced hybridization between particle and hole sectors. At sufficiently large anisotropy, the system also develops gapless regions in the Chern diagram where the topological invariant ceases to be well defined. Nevertheless, the momentum-space framework still predicts substantial parameter regions with finite band gaps and nontrivial Chern numbers. Within the conventional Thouless-pumping picture, such regions would ordinarily be interpreted as supporting robust quantized transport. As we demonstrate in the following section, however, the corresponding real-space dynamics reveal a fundamentally different behavior.

\subsection{Pumping path dependence}
Because experimental implementations inevitably involve imperfections in the pumping protocol, it is important to examine the robustness of the momentum-space topological characterization against temporal perturbations. To model deviations from the ideal pumping cycle, we introduce a higher-frequency perturbation to the staggered magnetic field,
\begin{equation} \label{eq:time_protocol_pert}
    \begin{split}
       &m(t)=\mathrm{sin}(\Omega t) + t_{\rm{p}}\ \mathrm{sgn}(\mathrm{sin}(\Omega t))\ \mathrm{sin}^{2}(f\Omega t)
    \end{split}
\end{equation}
where $t_p$ controls the perturbation strength and $f$ controls the perturbation frequency. 

\begin{figure*}
    \centering
    \includegraphics[width=1\linewidth]{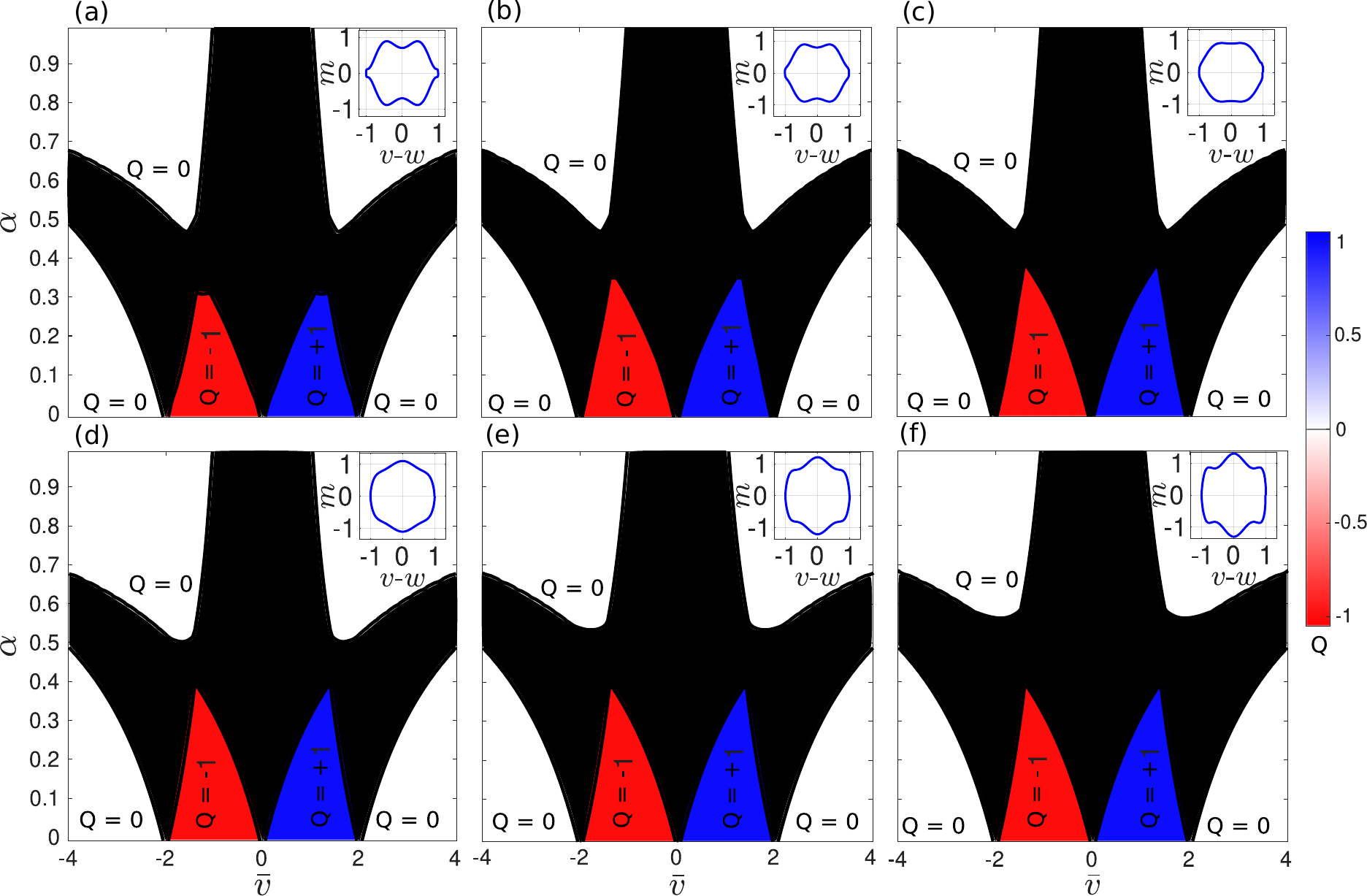}
    \caption{Effect of higher frequency temporal perturbation of pumping protocol on topological characterization of the system. The top row features the perturbation strengths $t_{\rm{p}} = -0.3\ \rm{(a)}, -0.2\ \rm{(b)}, -0.1\ \rm{(c)}$ respectively from left to right. The bottom row features the perturbation strengths $t_{\rm{p}} = 0.1\ \rm{(d)}, 0.2\ \rm{(e)}, 0.3\ \rm{(f)}$ respectively from left to right. The insets show the parameter space loops during one time period. The perturbation frequency is $f= 3$ (see \eqref{eq:time_protocol_pert}). In lower panels (d-f), i.e. for $t_{\rm{p}} > 0$ the region boundaries don't change, whereas in the upper panels (a-c), i.e. for $t_{\rm{p}} < 0$, the region boundaries show a very small shift.}
    \label{fig:phase_all}
\end{figure*}

Fig. \ref{fig:phase_all} shows the resulting Chern-number diagrams for some perturbation strengths $t_{\rm{p}}$ for a perturbation frequency $f = 3$. Results for different perturbation strengths are presented in different panels. The top row Fig \ref{fig:phase_all} (a-c) feature negative strengths and the bottom row Fig \ref{fig:phase_all} (d-f) feature positive ones. For sufficiently small perturbations, the overall topology of the phase diagram remains largely unchanged. In particular, parameter points located deep within the nontrivial regions preserve their quantized Chern numbers despite moderate distortions of the pumping protocol, although the topological characterization of the boundary regions appear to be affected. While this behavior, especially deep inside the non-trivial region, is consistent with the conventional expectation of topological robustness in momentum space, as we show next, this robustness of the Chern number does not necessarily imply robustness of the real-space spin transport itself in the anisotropic regime.  

\section{Real space dynamics}
The momentum-space analysis of the previous section predicts extended parameter regimes with nontrivial Chern numbers and finite band gaps, suggesting robust Thouless-type spin pumping. We now remove the periodicity condition and study the real-space time evolution under open boundary conditions. Our objective is twofold. First, we benchmark the isotropic limit, where the Rice–Mele correspondence is exact and quantized pumping is expected. Second, we investigate whether the same topological interpretation remains valid in the anisotropic regime, where pairing terms hybridize the particle and hole sectors. Throughout this section, we consider chains with $N = 10$ dimers and evolve localized end states under the pumping protocol introduced earlier.
\subsection{Quantized pumping in isotropic limit}
We begin with the isotropic limit of the chain ($\alpha \rightarrow 0$), where the Hamiltonian essentially separates into two independent Rice–Mele sectors. In this limit, the transported spin component along the z-axis is essentially conserved owing to the $SO(2)$ symmetry, and the momentum-space topology is expected to correspond directly to quantized real-space transport. At suitably chosen initial time the open chain supports localized end states at opposite ends of the system. These states form the two-dimensional pumping subspace responsible for transport during the pumping cycle.
We denote them as state 1 and state 2. 

\begin{figure*}
    \centering
\includegraphics[width=1\linewidth]{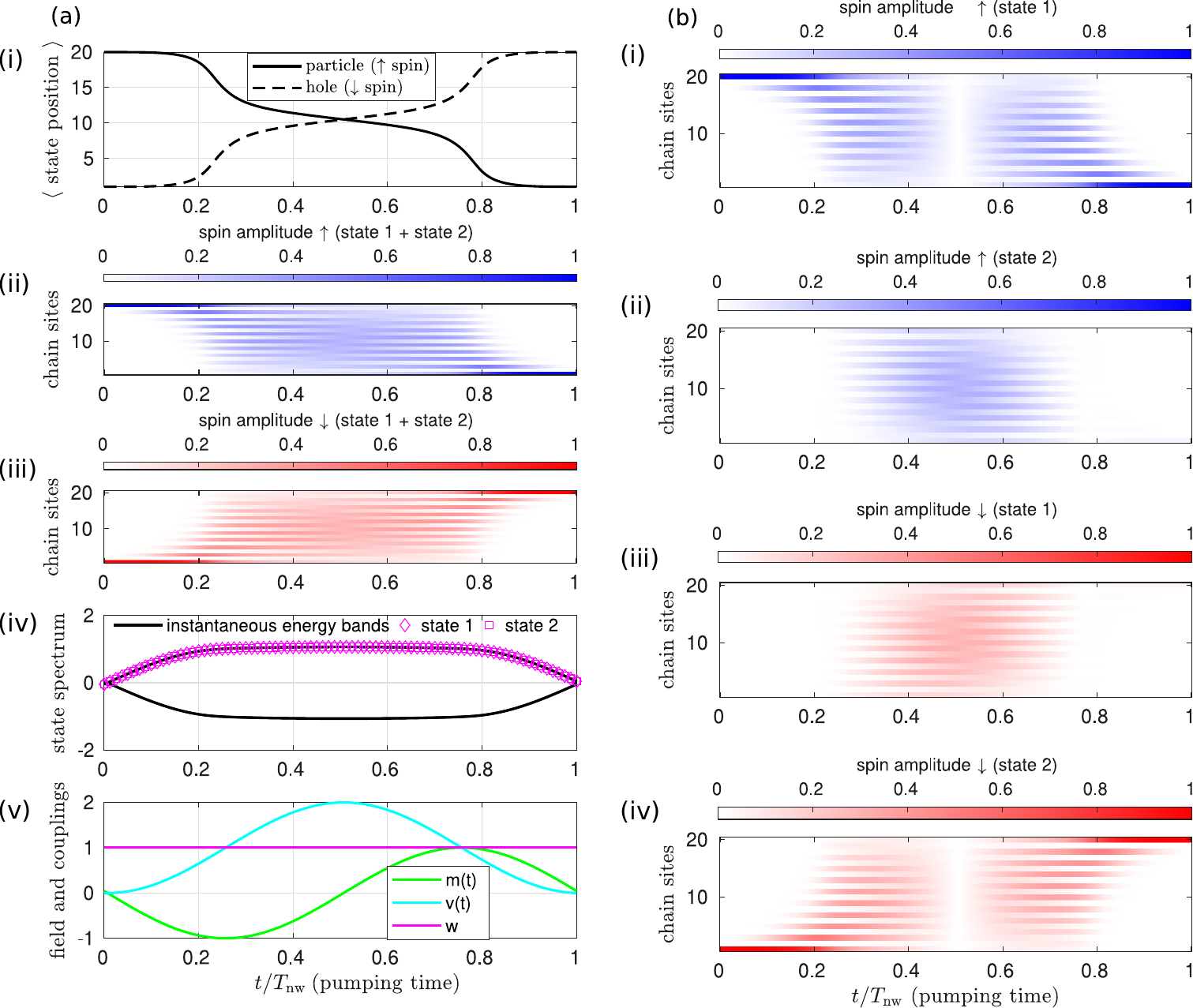}
    \caption{Real-space dynamics in the isotropic limit of the chain for $\bar v=1$. The pumping cycle transports particle-like and hole-like edge states between opposite ends of the chain while preserving quantized transport. All horizontal axes: pumping time. Colorbars: spin-amplitude. Panel (a-i) shows evolution of the expectation values for the particle ($\uparrow$-spin) and hole ($\downarrow$-spin) weights in states 1 and 2 combined. Panels (a-ii,iii) show time-evolution of particle ($\uparrow$-spin) and hole ($\downarrow$-spin) amplitudes over the entire chain. Panels (a-iv,v) show spectrum of the transporting states and temporal variation of system parameters. Panels (b) show the state resolved versions of (a-ii,iii).
    }
    \label{fig:iXYab}
\end{figure*}

Fig. \ref{fig:iXYab} shows the real-space evolution for a representative pumping cycle with $\bar v = 1$. Top three panels on the left show the expectation values of the position and total real-space distribution of particle- or hole-like weights of the states 1 and 2 over all the sites over the full time period of pumping. The bottom two panels on the left show the instantaneous energy bands (which coincide with state 1 and 2 at the initial time) and variation of the pumping parameters. The panels on the right show real-space distribution of particle- or hole-like weights in each of state 1 and 2. State 1 initially appears as a particle-like edge state localized near one end of the chain and evolves adiabatically to the opposite end during the pumping cycle. Simultaneously, state 2 behaves as a hole-like edge state propagating in the opposite direction. Each of the combined particle transport of states 1 and 2, and the combined hole transport of states 1 and 2 (measured by transfer of weights between the two endpoints of the chain) over a time period of pumping remains quantized and reproduces the momentum-space prediction $\mathcal Q=+1$. Importantly, the two transporting states remain essentially dynamically independent during the cycle. Because the Hamiltonian is essentially block diagonal in the isotropic limit in the fermionic representation, hybridization between the particle and hole sectors is almost negligible. During intermediate stages of the pumping cycle, the initially localized states partially delocalize into the bulk before re-localizing at the opposite edge.
For negative average intra-dimer coupling $\bar v < 0$, the direction of transport reverses while preserving quantization, consistent with the momentum-space characterization discussed previously. Thus the isotropic chain provides a consistent realization of topological pumping in both momentum-space and real-space descriptions.

\subsection{Loss of quantized pumping: anisotropic regime}
Next, we investigate the anisotropic regime, where the momentum-space analysis predicts nontrivial topological pumping over extended regions of parameter space despite the presence of pairing terms in the fermionic Hamiltonian. We focus on representative parameter values inside the topological region of the Chern diagram, such as $(\bar v,\alpha)=(0.8,0.2)$. According to the momentum-space characterization, this regime should support robust quantized pumping. The real-space evolution, however, reveals qualitatively different behavior. 

\begin{figure*}
    \centering
    \includegraphics[width=1\linewidth]{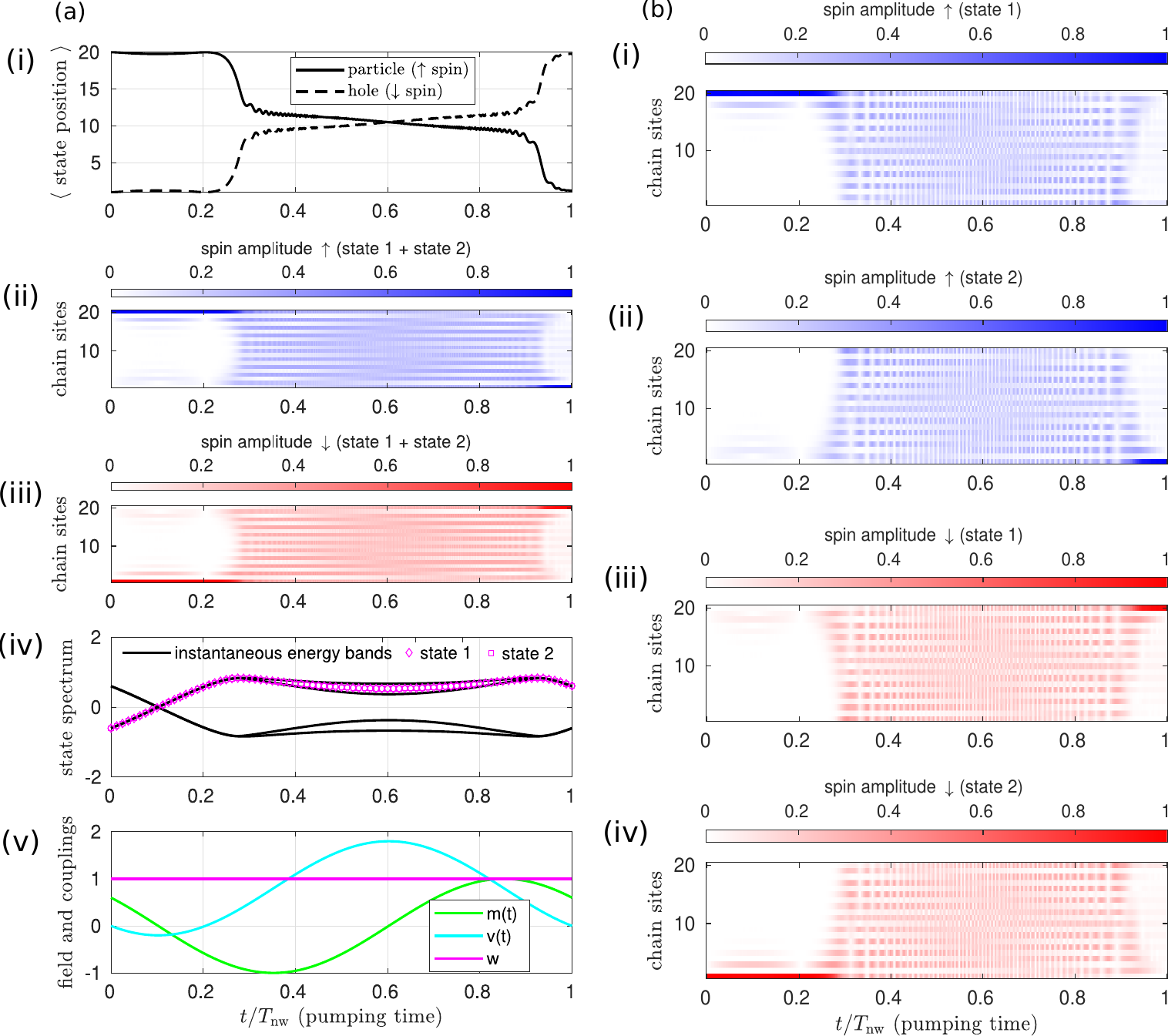}
    \caption{Real-space dynamics in the anisotropic regime for representative parameters inside the apparently topological region of the Chern diagram (yellow square, Fig. \ref{fig:PhaseDiag}). All horizontal axes: pumping time. Colorbars: spin-amplitude. Panel (a-i) shows evolution of the expectation values for the particle ($\uparrow$-spin) and hole ($\downarrow$-spin) weights in states 1 and 2 combined. Panels (a-ii,iii) show time-evolution of particle ($\uparrow$-spin) and hole ($\uparrow$-spin) amplitudes over the entire chain. Panels (a-iv,v) show spectrum of the transporting states and temporal variation of system parameters. Panels (b) show the state resolved versions of (a-ii,iii). Unlike the isotropic limit, the transporting states hybridize during the pumping cycle, resulting in non-quantized and protocol-dependent transport.
    }
    \label{fig:aXYab}
\end{figure*}

Fig. \ref{fig:aXYab} shows the transport dynamics in the anisotropic chain. Top three panels on the left show the expectation values of the position and total real-space distribution of particle- or hole-like weights of the states  1 and 2 over all the sites over the full time period of pumping. The bottom two panels on the left
show the instantaneous energy bands (which coincide
with state 1 and 2 at the initial time) and variation of
the pumping parameters as functions of pumping time. The panels on the right show real-space distribution of particle- or hole-like weights in each of state 1 and 2 as functions of pumping time. At early times, the evolution initially resembles the isotropic case: localized edge states begin propagating across the chain under the pumping protocol. As the cycle progresses, however, the transporting states increasingly hybridize with one another while spreading through the bulk (panels a-ii,iii and b). This behavior becomes particularly evident when the states occupy the chain non-locally during intermediate stages of the cycle. Unlike in the isotropic limit, the evolving states no longer remain confined to independent particle and hole sectors. Instead, the anisotropy-induced pairing terms dynamically mix the two sectors, causing the transporting states to become nontrivial superpositions of instantaneous eigenstates. Consequently, the transport no longer remains quantized. Although the momentum-space Chern number remains nonzero, the real-space pumped spin becomes sensitive to details of the evolution and loses the robustness expected of genuine Thouless pumping.

\begin{figure}
    \centering
    \includegraphics[width=1\linewidth]{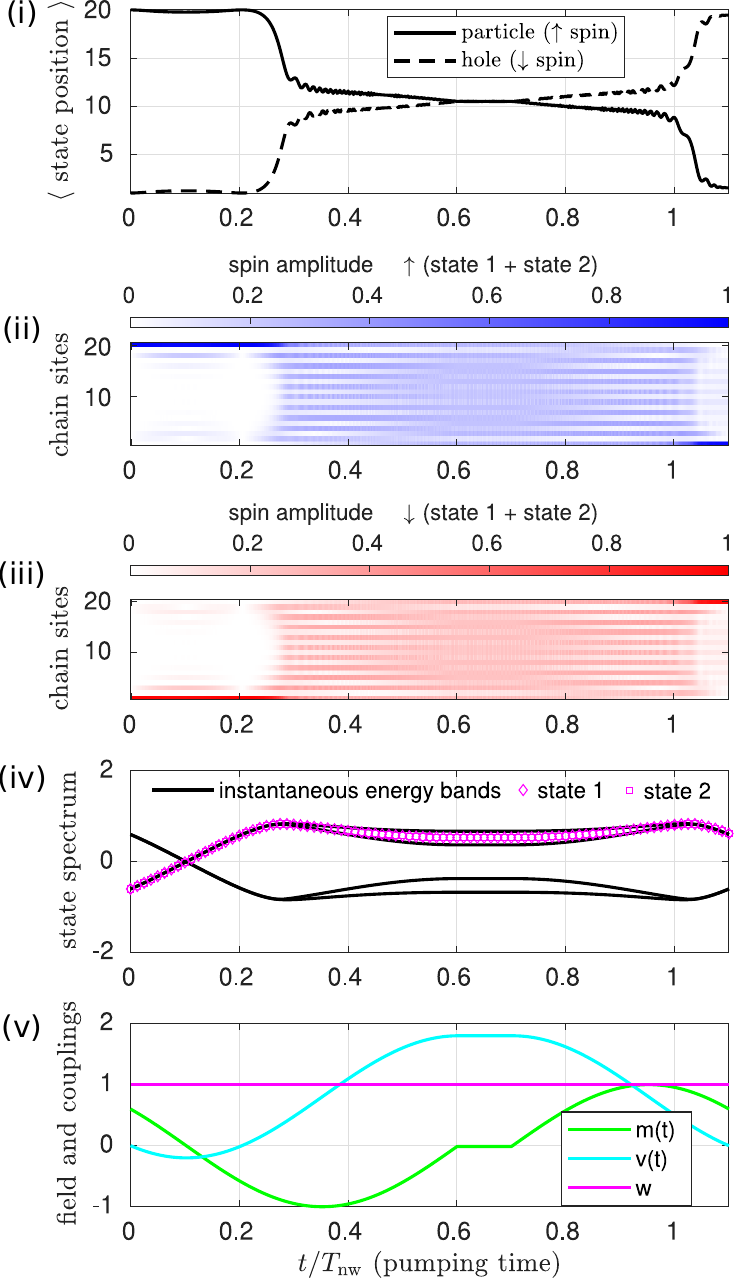}
    \caption{Representative wait-time pumping protocol in the anisotropic regime with waiting interval $t_{\rm{w}}=T_{\rm{nw}}/10$. All horizontal axes: pumping time. Colorbars: spin-amplitude. Panel (a-i) shows evolution of the expectation values for the particle ($\uparrow$-spin) and hole ($\downarrow$-spin) weights in states 1 and 2 combined. Panels (a-ii,iii) show time-evolution of particle ($\uparrow$-spin) and hole ($\downarrow$-spin) amplitudes over the entire chain. Panels (a-iv,v) show spectrum of the transporting states and temporal variation of system parameters.}
    \label{fig:nonzerowait}
\end{figure}
\subsection{Wait-time effect: quantization loss elaborated}
\begin{figure}
    \centering
    \includegraphics[width=1\linewidth]{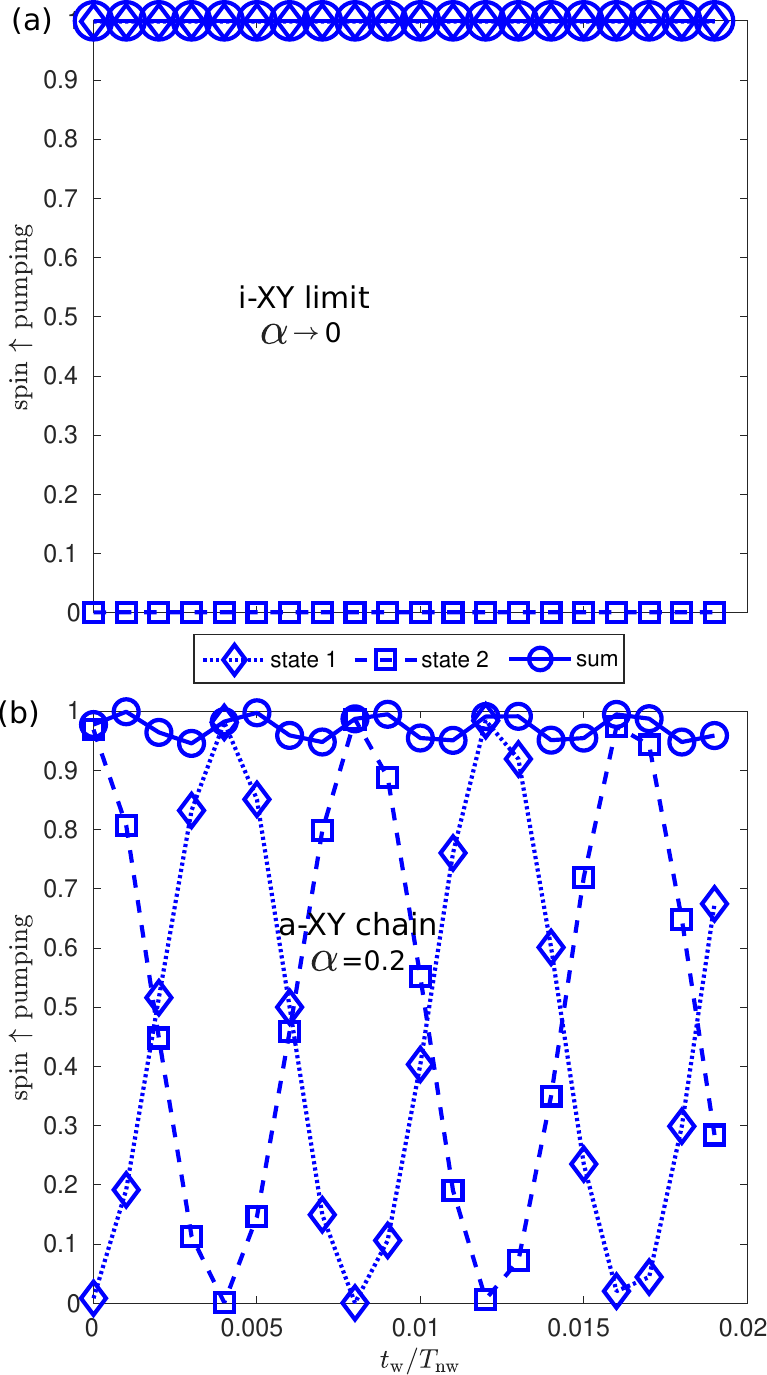}
    \caption{Relative transport contributions of the two pumping states as a function of waiting time $t_{\rm{w}}$. (a) Isotropic chain: the total transported spin remains quantized and independent of waiting time. (b) Anisotropic chain: the transported spin oscillates continuously with waiting time, and the sum also deviates from quantized value, demonstrating the absence of topological protection.
    }
    \label{fig:waittime_illus}
\end{figure}

To expose the non-protected nature of the transport more clearly, we next introduce a modified pumping protocol containing controlled waiting intervals.
The hybridization between the transporting states becomes especially transparent when the pumping cycle is modified to include finite waiting intervals during which the Hamiltonian is held fixed. In this wait-time protocol, the temporal variation of the pumping parameters is paused for a duration $t_{\rm{w}}$, after which the pumping cycle is resumed. During the waiting interval, the transporting states cease propagating spatially but continue evolving while having same energy. Fig. \ref{fig:nonzerowait} illustrates an example wait-time protocol with $t_{\rm{w}} = T_{\rm{nw}}/10$. Top three panels show the expectation values of the position and total real-space distribution of particle- or hole-like weights of the states. The bottom two panels show the instantaneous energy bands (which coincide with state 1
and 2 at the initial time) and variation of the pumping
parameters as functions of pumping time. During the paused interval, the localized edge states remain spatially frozen while oscillating between different superpositions within the pumping subspace. As a result, the relative contributions of states 1 and 2 to the final transported spin become strongly dependent on the waiting time. Fig. \ref{fig:waittime_illus} summarizes this behavior.

For the isotropic chain, shown in Fig. \ref{fig:waittime_illus} (a), the total transported spin remains exactly quantized for all waiting times. Although the phase evolution of the states changes during the waiting interval, the absence of any significant hybridization between particle and hole sectors preserves the quantized transport.
The anisotropic chain behaves fundamentally differently. As shown in Fig. \ref{fig:waittime_illus} (b), the transported spin oscillates continuously as a function of the waiting time and deviates substantially from integer values. Different waiting intervals therefore correspond to physically distinct transport outcomes in terms of state 1 and 2, despite sharing the same momentum-space topological characterization. 
This protocol dependence demonstrates explicitly that the transport in the anisotropic regime is not topologically protected. The momentum-space Chern number survives, but the real-space transport fails to exhibit the defining robustness of Thouless pumping. 
Thus, in the anisotropic regime, momentum-space topology mimics a spurious form of conventional topological pumping, while the corresponding real-space spin transport remains non-quantized and protocol dependent.

\section{Quantization loss: physical origin}
Our results reveal a marked distinction between the isotropic and anisotropic spin chains. In the isotropic limit, momentum-space topology and real-space transport remain consistent: nontrivial Chern numbers correspond directly to quantized and protocol-independent spin pumping. In the anisotropic regime, however, prediction of the same momentum-space characterization does not show real space quantized transport. The origin of this discrepancy lies in the spin conservation, or lack of it.

In the isotropic limit ($\alpha \rightarrow 0$), the X and Y directions are indistinguishable, and the Z component of the spin is conserved. The fermionized Hamiltonian separates into two independent blocks corresponding to particle and hole sectors. The pumping process can be mapped directly onto two decoupled Rice–Mele systems, and the conventional Thouless-pumping framework applies without modification. The anisotropic chain behaves fundamentally differently. For nonzero anisotropy, the fermionic Hamiltonian develops off-diagonal pairing terms that couple the particle and hole sectors dynamically during the evolution. The X and the Y directions are no longer equivalent, and the Z component of the spin is no longer conserved. This distinction is particularly important because the conventional Thouless-pumping argument implicitly assumes that the transported quantity is conserved throughout the adiabatic evolution.
{The wait-time protocol corroborates this cleanly, because a pause in the drive lengthens the cycle without altering the path traced in parameter space and therefore without altering any momentum-space invariant.}
The key point is that the momentum-space topology itself does not disappear in the anisotropic regime. The occupied bands continue to exhibit nontrivial Chern numbers over extended parameter regions. What fails instead is the conventional identification of these invariants with physically quantized spin transport.

\section{Discussion and outlook}
{Our results highlight an important distinction between the pumping captured in the momentum space topology and the real space transport. A nonzero Chern number can remain well defined and quantized, while the pumped states do not guarantee quantized transport from one end of the chain to the other end. This distinction becomes particularly important in presence of anisotropy. In the isotropic limit, conservation of $S^{z}$ and the separation into independent Rice--Mele sectors make these two notions coincide, allowing the Chern number to directly characterize state-resolved spin transport. Anisotropy removes this correspondence: particle--hole mixing makes the decomposition of the conserved subspace-level quantity among individual states dynamical and protocol dependent. The resulting distinction between topological pumping and state-resolved transport shows that a quantized band invariant alone need not imply quantized transport of the experimentally prepared state when the corresponding observable is not conserved. In this sense, the evolution of initially localized states from lower- to higher-energy bands while their end-to-end transport remains protocol dependent provides a concrete manifestation of the distinction between ``pumping'' inferred from band topology and physical real-space transport.}
{One could experimentally test the transport related predictions of our work in several systems, and compare those experimental transport results with the momentum space Chern numbers calculated for the same systems. This could provide a test of the discrepancy between the momentum space topology and the real space transport, that our work suggests.}

{In cold-atom chains the effective couplings and the staggered potential can be modulated with laser fields, giving access to a range of $\alpha$ and $\bar{v}$ across the phase diagram of Fig.~\ref{fig:PhaseDiag} and to the driving protocols used here; single-state preparation together with site-resolved readout makes the state-resolved transport directly observable, which is the quantity the present analysis identifies as
unprotected. In molecular magnetic complexes the exchange interactions are set mainly by synthesis, and dynamic tuning remains difficult, but modulation of the on-site anisotropy of a higher-spin chain has been proposed as a route to an effective spin-$\frac{1}{2}$ model with tunable $v$ and $w$~\cite{hoffman}, which together with a staggered field would allow selected predictions to be probed. The Kitaev-chain realizations in InAs nanowires~\cite{Sau2012,Dvir2023} remain another playground, since our model maps onto a Kitaev chain, and the transported quantity in that scenario is charge in a superconductor, which is itself not conserved. The same protocol-dependence of state-resolved transport should therefore appear for charge pumping in those devices.
{In these transport experiments, one would either need to make the systems isotropic, or have to carefully choose the pause-time in the pumping protocol to observe the intended real space transport dictated by the band topology.}}
\section*{Acknowledgments}
This work was supported as
part of the Center for Molecular Magnetic Quantum
Materials, an Energy Frontier Research Center funded
by the U.S. Department of Energy, Office of Science,
Basic Energy Sciences under Award No. DE-SC0019330.
Computations were done using the utilities of National
Energy Research Scientific Computing Center, and
University of Florida Research Computing systems. The author acknowledges useful discussions with Prof. Hai-Ping Cheng, Prof. Xiaoguang Zhang and Prof. Silas Hoffman.
\appendix  
\section{Pumping in the Rice-Mele model} 
\label{appendix:b}
\begin{figure*}[tb]
\centering
\includegraphics[width=1\linewidth]{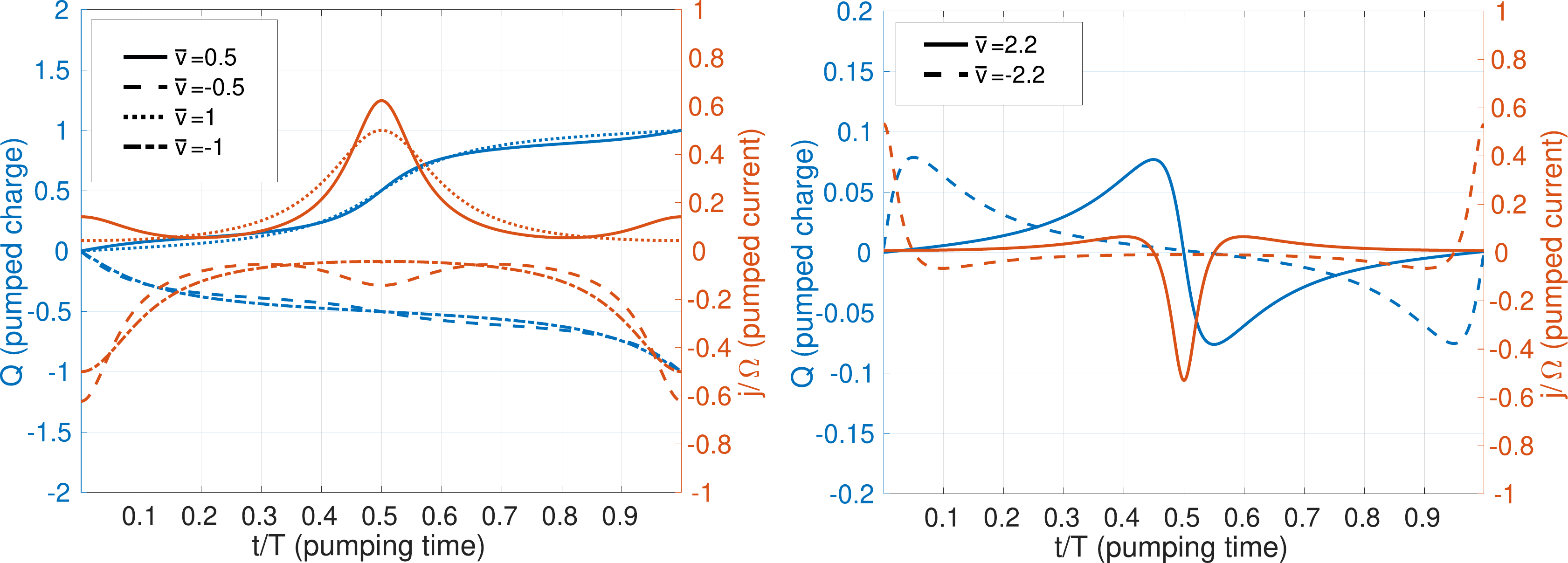}
\caption{Pumped current and charge in the Rice-Mele model over one time-period for various values of average intra-dimer hopping $\bar{v}$. Small magnitude of $\bar{v}$ gives non-zero pumping and the system is in a topological phase. Larger magnitudes drives the system into non-topological regime.}
\label{fig:RiceMeleCurrentCharge}
\end{figure*}

Here we present a brief recapitulation as well as reproduction of the pumped current and charge for various values of the intra-dimer coupling $\bar{v}$ in the simple Rice-Mele model, which are shown in Fig. \ref{fig:RiceMeleCurrentCharge}. Note that the Hamiltonian for the original Rice-Mele model is 2 $\times$ 2 (unlike the 4 $\times$ 4 fermionized Hamiltonian of the XY spin-chain) accommodating one filled band and no notion of in-plane anisotropy. The pumped charges demonstrate, as we see in the figure, discrete integer values within negligible numerical deviations. Small magnitudes of $\bar{v}$ pumps non-zero charge, thereby driving the system into topologically non-trivial phase. Calculated currents and charges for some typical low values of $\bar{v}$ corresponding to this phase is shown in Fig. \ref{fig:RiceMeleCurrentCharge}(a).
High magnitudes of $\bar{v}$ implies a strongly dimerized chain that supports no topological pumping in the system. As the magnitude of $\bar{v}$ increases gradually from smaller values, the pumped charge drops to zero, and the system enters a topologically {trivial} phase.
Results for some typical high values of $\bar{v}$ corresponding to this phase is shown in Fig. \ref{fig:RiceMeleCurrentCharge}(b).
\bibliography{References3.bib}
\end{document}